\documentstyle[twocolumn,prl,aps]{revtex}

\begin{document}

\draft
\title{Pulse-coupled relaxation oscillators: from
biological\\ synchronization to
Self-Organized Criticality}  \author{S.
Bottani\cite{email}} \address{Laboratoire de
Physique Th\'eorique et Hautes Energies, \\
 Universit\'{e} Paris VI\&VII,T24-5e, 2
place Jussieu, 75251 Paris Cedex 05,
France}

\date{\today}
\maketitle
\begin{abstract}It is shown that globally-coupled
oscillators with pulse interaction can synchronize under
broader conditions than widely believed from a theorem
of Mirollo \& Strogatz \cite{MirolloII}. This behavior
is stable against frozen disorder.  Beside the
relevance to biology, it is argued that synchronization
in relaxation oscillator models is related to
Self-Organized Criticality in Stick-Slip-like models.
(LPTHE preprint: 94/29)
\end{abstract}

\narrowtext
Large assemblies of oscillator units can
spontaneously evolve to a state of large scale
organization. Collective synchronization is the best
known phenomenon of this kind, where after some
transient regime, a coherent oscillatory activity of
the set of oscillators emerges. This effect has
attracted much interest in biology for the study of
large scale rhythms in populations of interacting
elements\cite{MirolloII}. The south-eastern fireflies, where
thousands of individuals gathered on trees flash
altogether in unison, is the most cited example
\cite{BuckI}. Other examples are the
rhythmic activity of cells of the heart pacemaker, of
cells of the pancreas, and of neural networks
\cite{BuckI,PeskinI,ErmentroutII}. Most of the works on
synchronization have used models in which the
interaction between the oscillators is smooth and
continuous in time (see e.g.
\cite{StrogatzIII}). Comparatively, few
analytical results are known on models where the
interactions are episodic and pulse-like
\cite{MirolloII,PeskinI,Gerstner,Tsodyks} although they
are relevant to several biological
situations as  fireflies and neural networks.
This letter deals with the emergence of
synchronization in a very simple model of
globally coupled Integrate and Fire (IF) oscillators,
see Eq.(\ref{MeanField}) for the definition and
\cite{BuckI}. It is shown that
synchronization occurs under broad conditions on the
properties of the oscillators, thus generalizing a
theorem of Mirollo and Strogatz
\cite{MirolloII}, the usual interpretation of which
restricted the synchronization conditions in a too
drastic way. Furthermore I show that the synchronized
state is stable against a frozen disorder of the
oscillator properties,  a subject much studied
in models with continuous interaction
\cite{StrogatzIII}. This result is different
from that of a previous study
\cite{Tsodyks} on a closely related model.

Beside synchronization another form of collective
organized behavior is known to occur in large
assemblies of elements with pulse interaction, that
is Self-Organized Criticality (SOC). This concept has
been proposed in  \cite{BakI} to describe out of
equilibrium systems that are generically critical,
i.e. that  organize into a scale invariant critical
state spontaneously, i.e. without  tuning of a
control parameter.
Systems displaying SOC are externally driven with a
drive  slower than any other characteristic time.
These models are made critical by the choice of a
threshold dynamics that forbids them to follow
adiabatically the external drive.
They can be
modelled as  coupled map
lattices, and can be divided into two subclasses
based on the concept of oscillators. In the subclass
(a) the external drive acts globally and
continuously on all the lattice sites of the
coupled map, until one of them reaches the
threshold, in which case it relaxes to zero: each
site is therefore a {\em relaxation oscillator}.
These are the stick-slip-like models \cite{OlamiI}.
They are deterministically \cite{OlamiI} or not
\cite{ChristensenI,Bottani} driven systems, for
which no conserved quantity is known, and can
exhibit SOC even when they are
dissipative\cite{OlamiI}. The systems of the
subclass (b) are driven at each time step by the
increment  of a unique site, so that sites are {\em
not} oscillators. These are the sandpile-like
models\cite{BakI,ZhangI}.
The model studied in this letter is a simple model
of fireflies flashing and is a mean field version
of class (a) models.
As this mean-field approach suggests, we argue in the
following that there is a close relationship between
SOC and
synchronization
\cite{ChristensenI,Bottani,Middleton}.

 The biological mechanisms controlling
the fireflies flashing have been
 extensively studied (see \cite{BuckI} for a
review) by testing the response of an isolated
insect to single or periodic flashes of light. It
appears that there are several  mechanisms for
synchronization and entrainment among the different
species, so that no general conclusion can be
drawn about both the mechanism and the biological
function of synchronized flashing. It is therefore
interesting in this context to study under which
general conditions a
synchronization may emerge in relaxation
oscillator models. Experiments indicate that
the rhythmic spontaneous flashing of the fireflies,
is in general governed by a flash-control pacemaker
in the brain, cycling as an oscillator with a given
intrinsic rhythm between a basal level of excitation
(state variable of the oscillator) and a fully excited
flash-triggering level
\cite{BuckI}. After each flash the state variable resets
quickly to the basal level. Light-flash stimuli on a
single firefly have different effects depending on the
species.  This has motivated the introduction of
essentially two models, the phase advance and phase
delay models
\cite{BuckI}. In the phase advance model, the
effect of  a pulse of light is to advance the
flashing of each insect pushing the state variable
towards the threshold. Flashing delay relative to
the intrinsic rhythm is impossible in this case.
This accounts for the behavior of at least two
species:  {\sl Photinus pyralis} and {\sl Photinus
concisus}, which however display only transient
synchrony possibly because these species are rover
\cite{BuckI}. On other species, as {\sl P.
cribellata}, a pulse of light inhibits the
oscillator, thus delaying the next flash. This is
interpreted in the phase delay model as the
resetting of the state variable to the basal level.
In this model, advance or delay of flashing is
possible. Let us mention that other fireflies species
have behaviors that cannot be described by any of
these two models.
The model we study in the
following \cite{MirolloII,ChristensenI} is equivalent
to the phase advance model.

\noindent {\em The mean field model.} It consists of
$N$ relaxation oscillators $O_i$ represented when
they evolve freely by a state variable
$E_i=E(t)\in[0,E_c=1]$ monotonically increasing in
time with  period $1$. The interaction is so defined
 that when
$E_i\ge1$ it relaxes to zero and increments all the
other oscillators by a pulse $\alpha/N$:
\begin{equation} E_i\ge 1 \Rightarrow
\left\{\begin{array}{lll}
E_i&\rightarrow& 0 \\
E_{i\neq j}&\rightarrow& E_j+{\alpha/N}
\end{array}\right.
\label{MeanField} \end{equation}
 where
$\alpha\in[0,1]$ is a dissipation parameter. The model
is taken into the slow drive limit  by assuming
that any  avalanche of firings, i.e. any succession of
firings triggered by the relaxation of one oscillator,
is instantaneous. We assume a supplementary rule that in
most circumstances does not change the behavior of the
system  but simplifies the discussion. This is the
so-called absorption rule of
\cite{MirolloII}, which is equivalent to a reaction
death time of the oscillators immediately after their
relaxation. This is taken into account by assuming
that oscillators  whithin the
same avalanche are not incremented by the pulses
resulting from the following relaxations in the
avalanche.

The theorem of Mirollo \& Strogatz \cite{MirolloII},
that applies to this model, states that for a convex
function $E_i=E(t)$ the system synchronizes
completely but for a set of initial conditions of
Lebesgue measure null. Although the validity of
this theorem cannot be questioned, we will see in
the following that for all the physically
interesting situations, the system synchronizes
even for a linear and a range of concave functions
$E(t)$.

Let us call configuration the set of ordered
distinct values $E_1^{(j)}< E_2^{(j)}<\dots<
E_{m_j}^{(j)}=1$ of the state variables present in the
system just before the $(j+1)$-th avalanche. To
each $E_i^{(j)}$ corresponds a group of $N_i^{(j)}$
oscillators at this value
$(\sum_{i=1}^{m_j}N_i^{(j)}=N)$. Let call
cycle, the time necessary for all the $m_j$ groups
to avalanche exactly once. To trace the evolution of
the system, it is useful to follow, cycle after
cycle, the gaps
$s_i^{(j)}=E_{i+1}^{(j)}-E_{i}^{(j)}$ between the
values of successive groups. If one of these gaps
$s_i^{(j)}$ becomes smaller than the value
$\alpha N_{i+1}^{(j)}/N$ of the pulse of the
$(i+1)$-th group, then the
$(i)$-th group gets absorbed by the $(i+1)$-th
group. For a linear $E(t)$, an elementary
calculation shows that the gap between a
group $(i+1)$ and a smaller one
$(i)$ is reduced during one cycle by $\delta
s_i=\vert\alpha(N_i-N_{i+1})/N\vert$.
Therefore large groups
unavoidably absorb the smaller ones that follow them
and become larger and larger \cite{Christensennote}:  it is this positive
feedback that leads to synchronization.
The only way for stopping this mechanism and therefore the evolution
towards synchronization is to get a configuration where
all groups are of equal size.  Apart in the random
initial conditions, where all the groups are of size
one, it is exceptional for large $N$  to be stuck
during the evolution in such a state. It
is in fact a difficult task to calculate in general
the probability for this situation to occur, but,
first, extensive numerical calculations show that
this is indeed exceptional even for
small $N$,  second, it is even a physically
ill-defined question since it depends on the
factorization of $N$, and third, if one imagines
that the oscillators are not exactly  identical,
the slightest splitting between their pulse
intensities forbids any $\delta s_i$ to vanish.
 Therefore to see if the system does not get stuck
in the initial  configuration,
the only relevant question is to know if it is
possible with random initial conditions to get
groups larger than one during the first cycle.
The
probability for pair formation may easily be
computed from the Poissonian distribution
$P(s)ds =Ne^{-Ns}ds,(N\gg1)$, of gaps $s$ between the
$N$ random numbers $E_i^{(1)}$ in $[0,1]$. The
number of oscillator pairs in the random initial
values satisfying the synchronizing conditions
$s<\alpha/N$ is then: \begin{equation}
N\int_0^{\alpha/N}P(s)ds=N(1-e^{-\alpha}), \qquad
N\gg1\label{Combien} \end{equation}  The absorption
process of oscillators into groups  takes place as
long as there is at least one such a gap. It
follows  directly from (\ref{Combien})  that this
is typically the case when $\alpha \ge 1/N$
\cite{ChristensenI}. Initial configurations of values
where no gap is smaller than
$\alpha/N$ are in principle possible, but for large
systems with a
$O(e^{-\alpha N})$ probability of occurence. In
conclusion, the set of linear oscillators completely
synchronizes almost always, when
$\alpha$ is of order $1$ compared to
$N$. The reason why this is not in contradiction
with the theorem of Mirollo \& Strogatz, is that
their recursive demonstration requires that two
single oscillators synchronize. While this is
 the case for convex oscillators, it is not
so for linear and concave ones. However, I disagree
with the ususal interpretation of the theorem of
Mirollo
\& Strogatz as the impossibility of
synchronization for non convex oscillators: in fact,
this theorem tells nothing in this case. In
\cite{Gerstner}, it was noted  that
convexity is not necessary for synchronization in the
case of a model  with transmission delays. We see here
that this condition is not even necessary.
Moreover, even concave oscillators can synchronize, provided that
the concavity is not too large. In this case, positive
feedback is in competition with concavity, that has for
effect to increase the gap between groups as they
approach the threshold. For small concavities positive feedback
prevails, whereas for larger concavity, groups are not able to grow.
This process can be seen on the class of concave functions $E(t)=t^a,a>1$.
Let us call $x_c=1-(1-n\alpha/N)^{(1/a)}$ the value of the phase
difference between two successive groups of $n$ and $m$
oscillators $(n>m)$ below which the first group absorbs the second one.
The first return map of the phase
difference $x_i$ between these two groups when the largest one gets to the
threshold (i.e. the state value of the second group is
$(1-x_i)^a$) has an attractive fixed point $x_0$.
The condition for synchronization is  $x_0<x_c$.
For a fixed pulse $\alpha/N$ there is an upper value for the
concavity $a_{max}=1-\log(1-\alpha/N)/\log 2$ below
which the preceding condition is fulfilled for all
group sizes $n>m$. In this case the system synchronizes
as with linear oscillators. Since $a_{max}$ is
the most stringent bound, we expect that synchronization
almost always occurs for a larger range of  values of $a$.
This has been extensively  verified   numerically \cite{Bottani}.

Up to now, all the oscillators
were identical. However it is known that firefly
populations have a spread of individual
frequencies \cite{BuckI}. Therefore it is necessary to
show that the phase delay model  accounts for
synchronization even in this case.

\noindent{\em Frequency distribution.}  Different
frequencies may be allowed in  IF models, by keeping
an identical threshold for the oscillators, but
letting them have different slopes
$E_i(t)=a_it\,,\: E_i\in[0,1]$, so that the $a_i$'s
are  the internal frequencies. Positive feedback is
here also the effective mechanism leading to
synchronization for small disorder of frequencies.
For larger frequency dispersion the formation of
large groups is not always possible, since two
oscillators relaxing together at one moment do not
necessarily relax again together at the next cycle.
In fact two questions must be  answered: first, can
stable groups exist, and second, can such groups be
formed during the evolution? As we will see, this
second point is not problematic, and stability
is the most stringent condition for the emergence of
synchronization. Let us first notice that a
necessary condition for the stability of a group of
$n$ oscillators, is that the relaxation of the
fastest one  triggers the avalanche of all the
others. Therefore stable groups must fire at the
rhythm of the elements with highest frequency. Let
us now study the stability of a group of $(n+1)$
elements in the background of the
 other $N-(n+1)$ oscillators, assuming that
the subset of the  $n$ first ones, $O_1,\dots,O_n$
of frequencies $a_1>\dots>a_n$ is stable, and that
the $(n+1)$-th oscillator $O_{n+1}$, has the lowest
frequency $a_{n+1}$. Assuming that $O_1,\dots
,O_n,O_{n+1}$ have just relaxed together, the
condition for again relaxing together  after a cycle
is:
\begin{equation}
1-a_{n+1}\left(\frac{1}{a_1}+\Delta\frac{a_1-a_{n+1}
}{a_1a_{n+1}}\right)\le
n\frac{\alpha}{N},
\label{ConditionStabl}
\end{equation} where $\Delta$ is the summed
strength of  the pulses of all the other
oscillators. Therefore the most stringent condition
of stability for a group of
$n+1$ oscillators is
\begin{equation}
\frac{a_1-a_i}{a_1}\le(i-1)\frac{\alpha}{N}\qquad
\forall i=1\dots n+1
\label{StringentCond}
\end{equation} Therefore, given a system of $N$
oscillators with a random uniform distribution of
frequencies centered on $a$ with width
$D=a_{max}-a_{min}$, the probability that the whole
system remains stable is:
\begin{eqnarray}
P_N(D/a,\alpha)&=&\rho^N\int_0^{\delta}
da'_1\int_{a'_1}^{2\delta}da'_2\dots
\int_{a'_{N-1}}^{N\delta}da'_N  e^{-\rho
a'_N}\nonumber\\ &=&1-e^{-\rho\delta}-\rho\delta
e^{-2\rho\delta}\nonumber\\
 & &\mbox{}-e^{-\rho\delta}\sum_{j=2}^{N-1}{(j+1)^{j-1}\over
j!}\left({\rho\delta} e^{-\rho\delta}
\right)^j.
\label{ProbaStab}
\end{eqnarray} where
$\rho=N/D$ and $\delta=\alpha a_{max}/N$. One can
see on Fig. \ref{fig1} that even for large $D/a$ there
is a relatively high probability to have a
realization of frequencies allowing complete
synchronization as long as $\alpha$ is not too
small, in which case the system decouples and
synchronization is less probable. One can verify
that, when the probability $P_N$ is not vanishingly
small, it is almost independent of $N$ for large $N$
$(\sim 100)$.  Now we will see that any two stable
groups of oscillators will, at some moment, relax
together. Although positive feedback is still
acting as in the case of identical oscillators, it
is in competition with disorder that tends to
desynchronize groups. However, contrary to the case
of identical linear oscillators, it is easy to
show that the dispersion of frequencies implies that
the relative positions of any two groups change
monotically in time. Therefore at some moment the
two groups must relax together. If their union is
itself  stable, it  forms a unique stable group.
This proves that if the distribution of frequencies
fulfills the stability conditions,
Eq.(\ref{StringentCond}), the system completely
synchronizes. If this condition is not satisfied,
there is partial synchronization:  the subsystems
that individually fulfill Eq.(\ref{StringentCond})
synchronize. Extensive numerical simulations confirm
all these conclusions \cite{Bottani}.

It is natural to think of other kinds of
frozen disorder. For instance, the pulses do not
necessarily need to be of the same strength, and
also the oscillators may have different threshold
values. As expected,
and for similar reasons as those
previously explained for the case of a
spread of frequencies,
synchronization occurs also in this case and
even for strong disorder. These results are different
from  the conclusions in \cite{Tsodyks} where for a
similar but different model, the completely
synchronized state is unstable even against weak
disorder.

\noindent{\em Lattice models.} Beside synchronization or
quasi-synchronization, lattice models of pulse-coupled
oscillators  may display  SOC. This is the case for the
models of class (a), for which however SOC seems to be
related to a tendency to synchronization
\cite{ChristensenI,Middleton}.
SOC also
appears when a system is perturbed, which otherwise
should synchronize totally or partially
\cite{ChristensenI,Bottani}, or which should be
periodic \cite{GrassbergerI}. The most striking
example of class (a) models is the
Olami-Feder-Christensen (OFC) model
\cite{OlamiI} that consists of oscillators $E_i$ on a
square lattice, that relax to zero when they exceed a
given threshold $E_c$, thus incrementing their
nearest neighbors by  a pulse which is
$\alpha\:(\alpha\le1/4)$ times their value:
\begin{equation} E_i\ge E_c \Rightarrow
\left\{\begin{array}{lll} E_i &\rightarrow &0 \\
E_{nn} &\rightarrow &E_{nn}+\alpha E_i
\end{array}\right. \label{OFC} \end{equation}
 With open boundary conditions this model
shows SOC, while the Feder
\& Feder (FF) model \cite{FederI}, which is
identical but for the increment, which is a constant
that can be seen as the mean of $\alpha E_i$ (pulse
of the FF model $=\Delta=\overline{\alpha E_i}$),
shows partial synchronization
\cite{ChristensenI,Bottani} (not SOC as claimed in
\cite{FederI}). This clearly means that the randomness
of the initial conditions, which is dyamically
eliminated in the FF model while it is mantained via
the increment in the OFC model, changes the behavior of
the system from partial synchronization to SOC.
Furthermore, different kinds of perturbations
incompatible with a periodic behavior, change
periodically ordered states for SOC.
 If, for instance, a random noise is added to the
increment in the FF model, synchronization
disappears and the system becomes SOC
\cite{ChristensenI}. On the other hand, if instead
of open boundary conditions, periodic
 conditions are used, SOC disappears in
favor of a periodic state, the period being the
number of sites of the lattice \cite{GrassbergerI}.
It reappears if one inhomogeneity is introduced on
the lattice \cite{GrassbergerI}.
Extensive numerical simulations \cite{Bottani} show
that  large avalanches are spatially
quasi-stable. This means that during  activation
cycles of the oscillators, sites participating in
large avalanches are synchronized. Thus the system finds
a compromise between synchronization and SOC although
this seems contradictory since SOC is incompatible
with complete synchronization \cite{Middleton,Bottani}.

In this letter I have shown that in the
mean-field, relaxation oscillator models
such as the IF phase advance model
synchronize under  not as constrained
conditions as suggested by the usual (erroneous)
interpretation of Mirollo \& Strogatz theorem.
Moreover, disorder, which is a constant of the real
world, does not spoil synchronization in this
model, opening  the prospect of a
larger applicability of IF oscillators models and of
variants.

I gratefully acknowledge B. Delamotte for his
advice and supervision, as well as D. Sornette for
fruitful discussions. I thank also C.
Tang for giving me a recent preprint with A.A.
Middleton developing  ideas
about SOC and synchronization  consistent with
mine \cite{Middleton}. Laboratoire de Physique
Th\'eorique et des Hautes Energies is a Unit\'e
associ\'ee au CNRS: D 0280.

\begin{figure}
\caption{ Plot of the probability $P_N$, Eq.($6$)
for
different dissipation levels.
 From top to bottom  $\alpha = 0.5, 0.4, 0.3, 0.2, 0.1$. }
\label{fig1}
\end{figure}

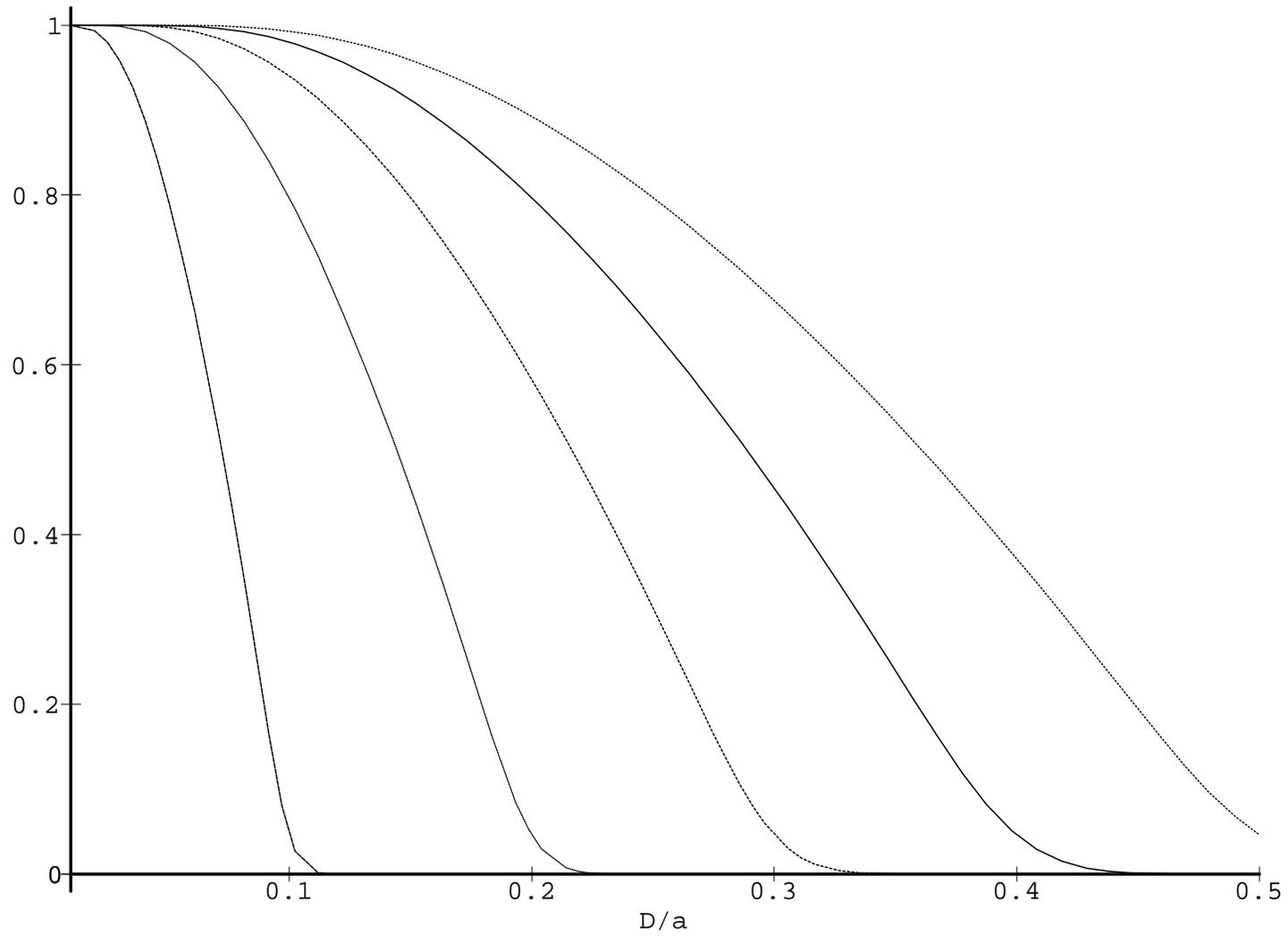
\end{document}